# Carbon nanotube synthesis from propane decomposition on a pre-treated Ni overlayer

J SENGUPTA, S K PANDA and C JACOB*

Materials Science Centre, Indian Institute of Technology, Kharagpur 721 302, India

**Abstract.** Growth of carbon nanotubes (CNTs) was performed by atmospheric pressure chemical vapour deposition (APCVD) of propane on Si(111) with a pre-treated Ni overlayer acting as a catalyst. Prior to the growth of CNTs, a thin film of Ni was deposited on Si(111) substrate by evaporation and heat treated at 900°C. The growth of nanotubes was carried out at 850°C using propane as a source of carbon. Distribution of the catalyst particles over the Si substrate was analysed before and after heat treatment by atomic force microscopy (AFM). The X-ray diffraction (XRD) pattern of the grown material revealed that they are graphitic in nature. Field emission scanning electron microscopy (FESEM) was used to investigate the growth process and it was found that a catalytic particle was always situated at the tip of the tube thus implying a tip growth mechanism. Evidence for the presence of radial breathing mode from multi-wall nanotubes (MWNTs) in the grown sample was obtained from micro-Raman analysis. Finally, high-resolution transmission electron microscopic (HRTEM) analysis confirmed that the graphene layers of the CNTs are well ordered with typical 0·34 nm spacing.

**Keywords.** Carbon nanotubes; Ni-catalysed growth; APCVD; AFM; HRTEM; micro Raman spectroscopy.

## 1. Introduction

An unique combination of electronic, thermal, mechanical and chemical properties (Dai *et al* 1996; Saito *et al* 1997; Tans *et al* 1998; Chen *et al* 1999; Qian *et al* 2000) have made CNTs a centre of attraction in the field of nanoscale research since their discovery in 1991 by Iijima (Iijima 1991). CNT-based field-emission devices exhibit outstanding emission characteristics (Heer *et al* 1995; Kim *et al* 2000) and remarkable current stability (Dean and Chamala 1999; Zhu *et al* 1999) as an outcome of the high aspect ratio, mechanical stiffness and electrical conductivity of the nanotubes (Wong *et al* 1997; Yao *et al* 2000). Due to the large surface area, low resistivity and chemical stability of nanotubes along with the capability to grow directly onto desired substrates, CNTs emerge as a potential electrode material for electrochemical energy storage devices and supercapacitors (Leroux *et al* 1999; Nutzenadel *et al* 1999; Frackowiak *et al* 2000).

Although many basic experiments are centred on SWNTs but such materials produce at high temperatures and are hard to manipulate into useful configurations. Conversely, MWNTs grow at lower temperatures directly onto substrates in useful geometries (Huang *et al* 1998; Ren *et al* 1998). Compared to SWNTs, the larger wall thicknesses of MWNTs lead to larger diameters, higher stiffness, and highly improved electrical conductivity; which make MWNTs a suitable candidate for cold-cathode electron field emission, nanoporous membranes, low-friction nanobearings, ultra-sensitive electrometers and other applications (Shiflett and Foley 1999; Cummings and Zettl 2000; Roschier *et al* 2001).

The synthesis of CNTs can be grouped into the following categories (Iijima 1993; Thess *et al* 1996; Journet *et al* 1997; Meyyappan *et al* 2003): arc discharge, chemical vapour deposition, plasma method, laser ablation, etc. Among the above methods, CVD method is simple and easy to implement, and has been widely used because of its potential advantage to produce a large amount of CNT growing directly on a desired substrate with high purity, large yield and controlled alignment, whereas the nanotubes must be collected separately in the other growth techniques (Iijima 1993; Thess *et al* 1996; Journet *et al* 1997; Kong *et al* 1998; Meyyappan *et al* 2003).

To achieve a controllable growth of the CNTs, an understanding of their growth mechanism is of importance. The general growth process of CNTs by CVD is based on the following mechanism proposed by Baker *et al* (1972) known as the vapour–liquid–solid (VLS) mechanism. In this mechanism, liquid catalytic particles at high temperature absorb carbon atoms from the vapour to form a metal–carbon solid state solution. When this solution becomes supersaturated, carbon precipitates at the surface of the particle in its stable form and lead to the formation of a carbon tube structure. Depending on the position of

*Author for correspondence (cxj14_holiday@yahoo.com)

the catalyst particle, base and tip growth mechanisms have been proposed (Tesner *et al* 1972; Oberlin *et al* 1976; Tibbets 1984).

The metal clusters acting as a catalyst for CNT growth by CVD can be produced by different methods, e.g. by vapour or sputter deposition of a thin metallic layer on the substrate, by saturation of porous materials by metals, by deposition of solutions of substances containing catalyzing metals or by introduction of organometallic substances into the reactor (Jeong *et al* 2002; Tu *et al* 2002; Huang *et al* 2003; Bertoni *et al* 2004). Prior to CNT synthesis, high temperature hydrogen treatment is an important step in order to produce contamination free catalyst particles and for the removal of the oxides that may exist over catalyst surface (Takagi *et al* 2007).

In this paper, we have discussed the formation of carbon nanotubes using APCVD over a pre-heated Ni catalyst on Si(111) substrate using propane as a source of carbon. AFM study was performed on the Ni catalyst, before and after heat treatment to observe the change in morphology due to heating. FESEM and HRTEM were used to characterize the growth morphology and structure of the nanotubes. Compositional information was obtained by XRD and micro-Raman spectroscopy.

## 2. Experimental

Atmospheric pressure chemical vapour deposition of CNTs was carried out by catalytic decomposition of propane on Si(111) wafers with a pre-treated Ni overlayer in a hot-wall horizontal CVD reactor using a resistance-heated furnace (ELECTROHEAT EN345T).

The Si (111) substrates were ultrasonically cleaned with acetone and deionized water prior to Ni film deposition. For Ni deposition, the sample was loaded in a vacuum system (Hind Hivac: Model 12A4D) and pumped down to a base pressure of $10^{-5}$ Torr and Ni thin films (thickness ~ 20 nm) were deposited by evaporation of metallic nickel (purity, ~ 99·994%).

The substrates were then loaded into a quartz tube furnace, pumped down to $10^{-2}$ Torr and backfilled with flowing argon. When the furnace temperature stabilized at 900°C, the samples were annealed in hydrogen atmosphere for 10 min. Finally, the reactor temperature was brought down to 850°C and the hydrogen was turned off, thereafter propane was introduced into the gas stream at a flow rate of 200 SCCM, for 1 h for CNT synthesis.

A nanonics Multiview 1000$^{TM}$ system in AFM mode was used to image the surface morphology of the Ni layer before and after heat treatment with a quartz optical fibre tip in tapping mode. Samples were also characterized by a Rigaku X-ray diffractometer (ULTIMA III) with CuK$\alpha$ source and $\theta$–$2\theta$ geometry to analyse the crystallinity and phases of grown species. Micro Raman measurements were carried out at room temperature in a backscattering geometry using a 488 nm air-cooled Ar$^+$ laser as an excitation source for compositional analysis. FESEM (ZEISS SUPRA 40) and HRTEM (JEOL JEM 2100) were employed for examination of the morphology and microstructure of the CNTs.

## 3. Results and discussion

### 3.1 *AFM analysis*

Figure 1 shows the 3-D AFM image of the distribution of the as deposited catalyst particle over the Si (111) substrate. The AFM image reveals that the initial film consists of Ni clusters instead of a continuous layer. As Ni atoms interact more strongly with each other than they do with the Si substrate hence the catalyst deposition proceeds via island nucleation and coalescence, in accor-dance with the Volmer–Weber mode (Tiller 1991). After deposition of Ni, the sample was annealed in hydrogen for 10 min at 900°C. This heat treatment results in the formation of Ni islands as confirmed by AFM image in figure 2. Heating above a certain temperature causes clusters to coalesce and form macroscopic islands. This process is based on cluster diffusion and depends on their density and diffusion constant, at a given substrate temperature. Cluster diffusion terminates when the island shape is of minimum energy for the specific annealing conditions (Jak et al 2000). These clusters act as a catalyst surface for nanotube growth.

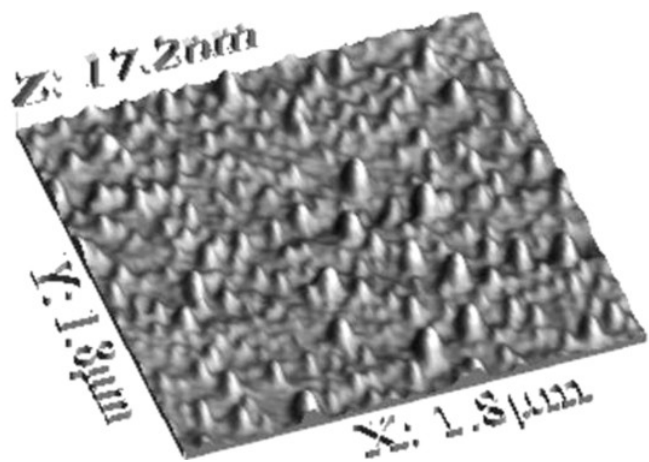

**Figure 1.** A 3D AFM image of the as-deposited Ni film, showing the clusters forming at the early stages of growth.

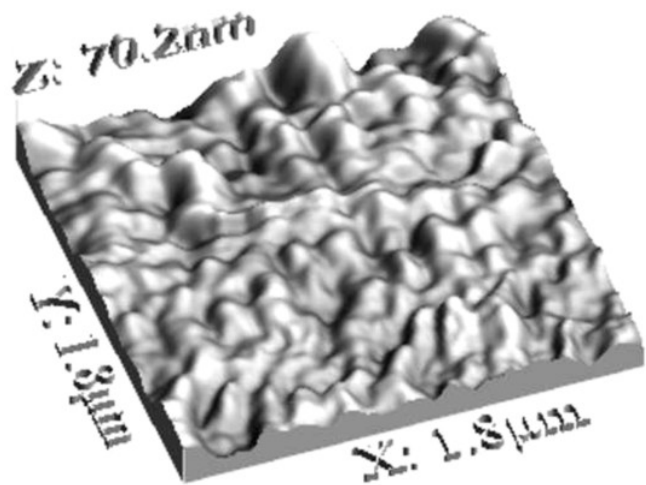

**Figure 2.** A 3D AFM image of Ni islands formed upon annealing the substrate at 900°C.

### 3.2 X-ray diffraction analysis

A XRD measurement was carried out using Cu K$\alpha$ radiation ($\lambda = 1.54059$ Å) to examine the structure of the CNTs and the resulting $\theta$–$2\theta$ scan is shown in figure 3. The peak at 26·3° is the characteristic graphitic peak arising due to the presence of multiwall carbon nanotubes (MWNTs) in the sample with inter-planar spacing of 0·34 nm (Cao et al 2001). The interlayer spacing of MWNT ($d$(002) = 0·344 nm) in the radial direction is similar to that of graphite ($d$(002) = 0·335 nm). The slightly larger $d$ spacing compared to that of graphite is related to the stacking disorder in these nanostructures. The peak near 43·6° is attributed to the (101) plane of the nanotube (Chatterjee et al 2003). The peaks at 28·5° and 58·9°, however, did not originate from the MWNTs and are attributed to (111) and (222) planes, respectively of the Si substrate.

### 3.3 FESEM analysis

Field emission scanning electron microscopy (FESEM) was employed for the analysis of the morphology and density of MWNTs. Figures 4a and b are the FESEM micrographs showing the surface morphology of Ni-coated Si wafers after propane exposure. High-aspect ratio nanostructures are observed on the Ni film surfaces; there is no observable growth of these structures on bare Si substrates. The area density of the deposited MWNTs was high and the MWNT structures had randomly oriented spaghetti-like morphology. We note that in figure 4a the number of small tubes is much higher than the big ones and the tube diameter is in the range of 15–100 nm. Figure 4b shows the high magnification image of the small tubes with a narrow diameter distribution around 20 nm. In many cases, small bright catalyst particles were detected at the tip of the tubes. This suggests that the tip growth mechanism is likely to be responsible for the MWNT synthesis under the present conditions.

### 3.4 HRTEM analysis

High-resolution transmission electron microscopy (HRTEM) was used to characterize the growth morphology and structure of the nanotubes. Figure 5a shows high-resolution transmission electron microscopic (HRTEM) images of a MWNT grown from the Ni nanoparticle. It can be clearly seen that the CNT is well graphitized with an inner diameter of about 14–17 nm and outer diameter, 38–42 nm. The thickness of the tube wall lies in the range of 12–14 nm, which suggests that the tube wall is composed of ~ 30–40 graphitic layers. In the same figure, various defects and bamboo-like structures (diaphragms) inside the CNT can also be observed. The distance between two neighbouring graphitic walls is about 0·34 nm (inset); in agreement with the interplanar distance of CNT ($d$002) of 0·344 nm as obtained from XRD. Figure 5b shows the selected area diffraction (SAD) pattern from the MWNT shown in figure 5a. There are diffuse halos due to the amorphous carbon film on the copper grid and sharp rings due to the MWNT. The diffraction rings corresponding to a layer distance of 0·34 and 0·17 nm, are considered to be of the graphite (002) and (004) diffractions, respectively. The diffraction rings corresponding to layer distances of 0·20 and 0·12 nm can also be seen in figure 5b. The former corresponds to (101) and the latter to (110) of graphite.

### 3.5 Raman analysis

Micro-Raman spectroscopy provides more details of the quality and structure of the materials produced. The Raman spectra were taken in the backscattering geometry on the sample as grown. Figure 6 shows the room temperature Raman spectra of the MWNT material at a laser excitation wavelength of 488 nm. The spectrum is divided into two main zones: the low frequency region from 150–800 cm$^{-1}$ (figure 6a) and the high frequency zone from 1200–1700 cm$^{-1}$ (figure 6b). The vibrations of CNT originate from the curvature induced strain due to misalignment of the π-orbitals of adjacent coupled carbon atoms. These vibrations are reflected in the Raman peaks.

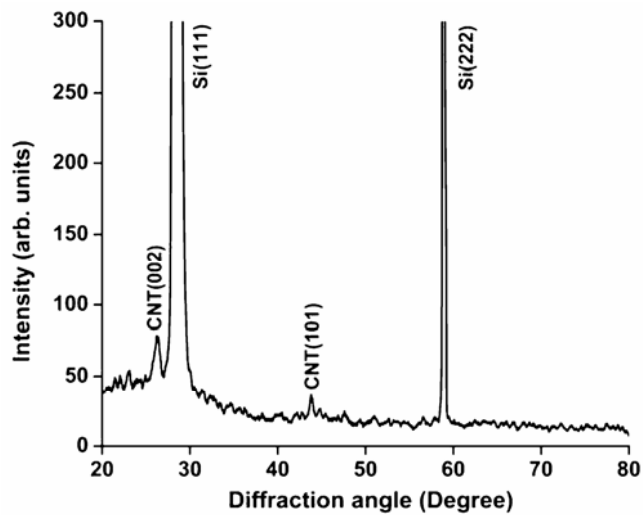

**Figure 3.** X-ray diffraction spectra of MWNTs grown on Si (111) substrate.

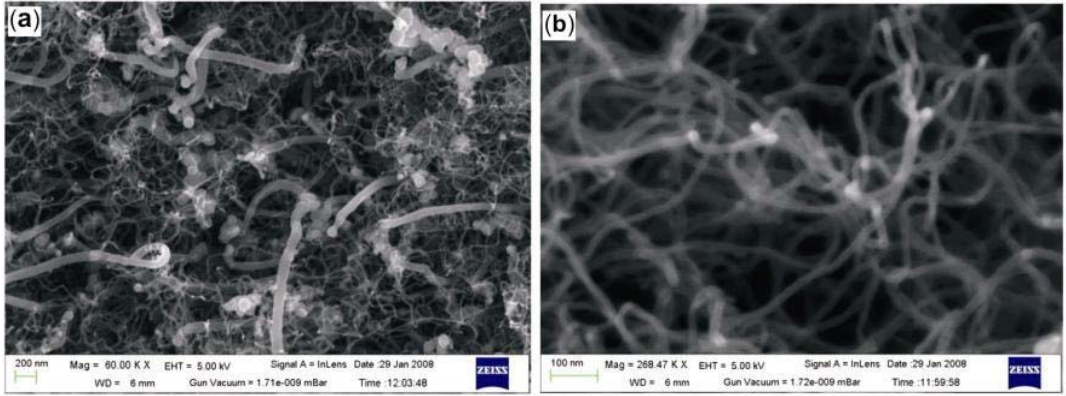

**Figure 4.** **a**. FESEM micrograph of the as-grown MWNTs deposited using the APCVD method and **b**. high magnification FESEM image of the small tubes.

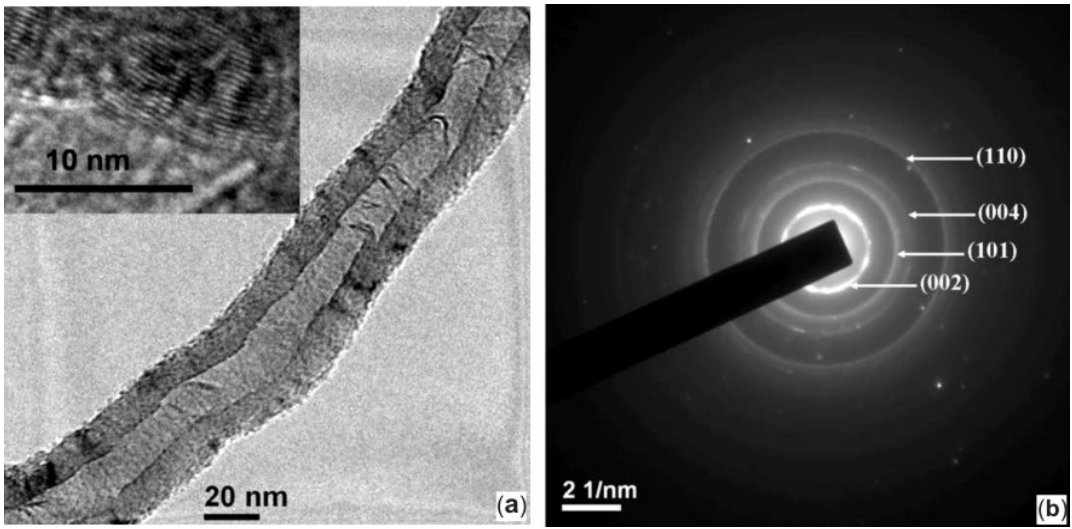

**Figure 5.** **a**. HRTEM image of the well-graphitized MWNT grown from the Ni using the APCVD method. (Inset) The fringe spacing between two carbon layers of MWNTs is about 0·34 nm and **b**. the selected area diffraction (SAD) pattern from the MWNT shown in figure 5a.

Evidence for the presence of radial breathing mode vibrations in this sample is obtained from the low wave number range of the spectrum (figure 6a). The peaks located at 220 cm$^{-1}$ and 286 cm$^{-1}$ are due to the radial breathing mode of MWNT (Jinno *et al* 2004). This mode has $A_{1g}$ symmetry and all the carbon atoms move in phase in the radial direction creating breathing like vibration of the entire tube (Zhao *et al* 2002). The frequency of RBM is directly linked to the innermost tube diameter by the relation

$\omega_{RBM} = 223 \cdot 75/d,$

where $d$ is the innermost tube diameter and $\omega_{RBM}$ the wave number in units of nm and cm$^{-1}$ (Jinno *et al* 2004).

The peaks at 220 cm$^{-1}$ and 286 cm$^{-1}$ originate due to the radial breathing mode of MWNT bundle with inner-core diameter of 1 nm and 0·78 nm, respectively.

The two main peaks observed in the high frequency zone (figure 6b) are the so-called *D*- and *G*-lines (Eklund *et al* 1995). The *G*-line appears at 1567 cm$^{-1}$ and corresponds to the $E_{2g}$ mode i.e. the stretching mode of the C–C bond in the graphite plane and demonstrates the presence of crystalline graphitic carbon.

The *D*-line, at 1350 cm$^{-1}$, originates from disorder in the $sp^2$-hybridized carbon and can be due to the presence of lattice defects in the graphite sheet that makes up carbon nanotubes. The position of the *D* band for MWNT can be expressed as $\omega_D = 1285 + 26 \cdot 5 E_{laser}$ (Wei *et al* 2003). In our case, $E_{laser}$ = 2·54 eV resulting in $\omega_D$ = 1352·3 cm$^{-1}$, which is in agreement with our observed peak position and the FWHM is 43 cm$^{-1}$, whereas Dillon *et al* (2004) have stated the full width at half-maximum (FWHM) of the *D*-band of amorphous carbon and nano-crystalline graphite at 488 nm excitation as ~ 57 and 86 cm$^{-1}$. On the basis of these two characteristics (frequency and linewidth), we can conclude that the presence of the *D* band in the sample is due to the MWNT. However, a recent Raman analysis of multiwall nanotubes (MWNT) suggests that the *D* band is an intrinsic feature of the Raman spectrum of MWNTs, and they are not necessarily an indication of a disordered wall structure (Rao *et al* 2000).

The degree of graphitization is an indicator of the carbon nanotubes disorder level, and is evaluated by the intensity ratio of the *D* to *G* peaks ($R = I_D/I_G$). The intensity ratio derived from figure 6b is $R = I_D/I_G = 0.2$, indicating that the grown MWNT is highly crystalline in nature.

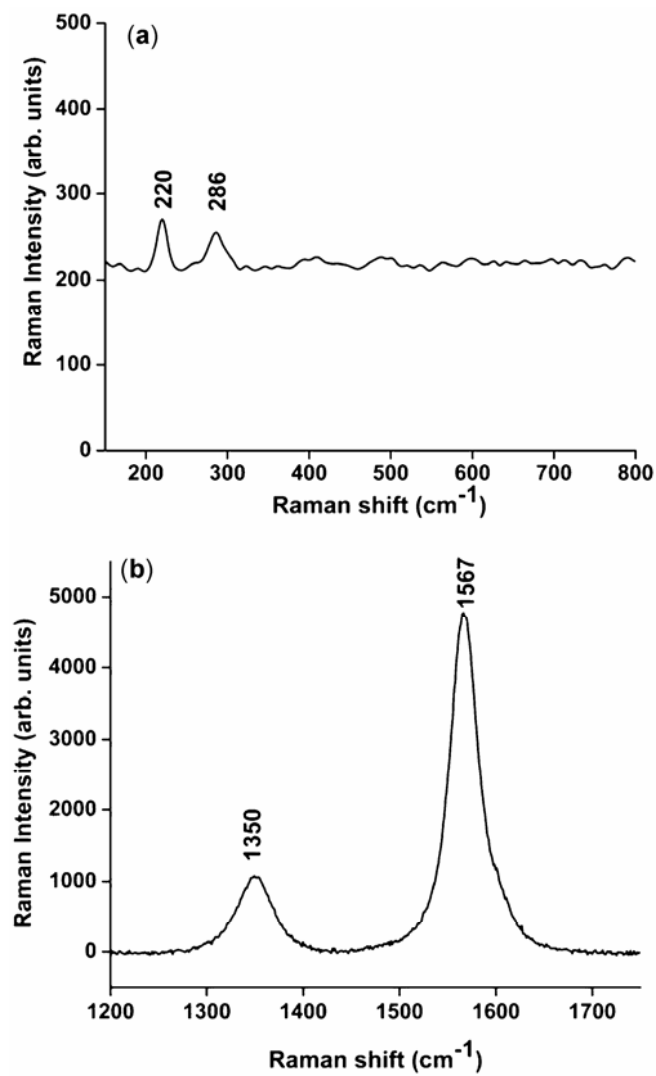

**Figure 6.** **a**. Low frequency Raman spectra (488 nm excitation) of a MWNT film grown by APCVD on Si showing the presence of RBM and **b**. high frequency Raman spectra (488 nm excitation) of a MWNT film grown by APCVD on Si.

## 4. Conclusions

We have grown MWNT by catalytic decomposition of propane on Si with a pre-treated Ni overlayer at 850°C by atmospheric chemical vapour deposition. At the growth temperature of 850°C, the catalyst particles coalesce on the substrate as confirmed by AFM studies. The XRD results suggest that MWNTs with good graphitization have been grown. Micro Raman analysis shows that radial breathing modes are present in the spectrum, generated from small diameter MWNTs. The FESEM images lead to the evidence that, in this experiment, MWNTs growth with Ni catalyst was primarily by the tip-growth mechanism. Lastly, HRTEM studies prove that the grown materials are carbon nanotubes and give a layer spacing of 0·34 nm. Since the catalyst particles are produced and deposited on a substrate using a dry process, the current method is applicable to any types of substrates. Moreover, many types of nanoparticles can be produced by simple evaporation. This will help in the optimization of catalyst materials for future studies.

## Acknowledgement

We are grateful to Dr A Roy, Department of Physics, Indian Institute of Technology Kharagpur, for her help in the Raman measurement.